# Global 3D hydrodynamic modeling of absorption in Lyα and He 10830 A lines at transits of GJ3470b

I. F. Shaikhislamov[1], M. L. Khodachenko[2,1,3], H. Lammer[2], A. G. Berezutsky[1,3], I. B. Miroshnichenko[1], M. S. Rumenskikh[1,3]

1) Institute of Laser Physics SB RAS, Novosibirsk, Russia
2) Space Research Institute, Austrian Academy of Sciences, Graz, Austria
3) Institute of Astronomy, Russian Academy of Sciences, Moscow, Russia
E-mail address: ildars@ngs.ru

**ABSTRACT**
Warm Neptune GJ3470b has been recently observed in $2^3S$-$2^3P$ transition of metastable helium, yielding absorption of about 1% in Doppler velocity range of [-40; 10] km/s. Along with previous detection of absorption in Lyα with depth of 20-40% in the blue and red wings of the line, it offers a complex target for simulation and testing of the current models. Obtained results suggest that absorption in both these lines comes from interaction of expanding upper planetary atmosphere with stellar plasma wind, allowing to constrain the stellar plasma parameters and the helium abundance in planet atmosphere.

**1. INTRODUCTION**

HST observations of Lyα flux at transits of close-in exoplanets revealed that a number of hot Jupiters and warm Neptunes are surrounded by extended hydrogen envelopes. The discovery and subsequent verification of 7-15 % absorption at transits of HD209458b (*Vidal-Madjar et al. 2003, 2004; Ehrenreich et al. 2008*) was followed by detection of Lyα absorption for HD189733b (*Lecavelier des Etangs et al. 2010, 2012*) and 55 Cnc b (*Ehrenreich et al. 2012*). The most prominent example of such absorption is the warm Neptune GJ 436b for which extremely deep transits up to 60% were measured (*Kulow et al. 2014; Ehrenreich et al. 2015; Lavie et al. 2017*). Moreover, this strong absorption takes place mostly in the blue wing of the line in the range of Doppler shifted velocities up to ~100 km/s, while the good signal-to-noise ratio revealed for the first time of observations in FUV such details of the transit light-curve as early ingress and extended egress phases with the total transit duration up to 25 hours.

*Salz et al. 2016* performed aeronomy hydrodynamic simulations of the escaping atmospheres of 18 hot gas planets in the solar neighborhood with the aim to find ideal observational targets. This study revealed that the warm Neptune GJ 3470b should have one of the largest mass loss rates due to its low mass ($M_p$=0.04$M_J$, $R_p$=0.37$R_J$,) close orbit ($R_{orb}$=0.0336 a.u) and relatively high activity of the host star, M1.5 dwarf ($M_{st}$=0.54$M_{Sun}$, $R_{st}$=0.55$R_{Sun}$). Observations within the frame of GTC survey (*Chen et al. 2017*) showed that the slope in optical transit spectrum is consistent with the Rayleigh scattering in an extended hydrogen-helium atmosphere. The first Lyα observation of GJ 3470b in three independent epochs, reported in (*Bourrier et al. 2018*), indeed revealed the large absorption depth of 35±7 % in the blue wing [-94; -41] km/s of the line. However, as a special feature of this particular planet, different from the similar warm Neptune GJ 436b, is the absorption at the level of 23±5 %, measured also in the red wing [23; 76] km/s, as well as a relatively short transit duration of ~2 hours without any distinct early ingress and extended egress phases.

Altogether, GJ 3470b appears to be the second hot exoplanet with a large hydrogen envelope extending far beyond the Roche lobe. Moreover, besides of the hydrogen related Lyα, also the absorption in helium line has been detected for this planet. Recently it was suggested (*Oklopcic & Hirata 2018*) that the absorption by a metastable helium in the $2^3S$ state at 10830 Å offers an alternative way to probe the evaporating exoplanetary atmospheres. It is not affected by the interstellar medium and can be observed by ground telescopes. *Ninan et al. 2019* reported about 1% absorption in three transits of GJ 3470b, calculated by averaging over the line width of 1.2 Å (or 33 km/s). Despite relatively low S/N ratio, the data indicate that absorption takes place mostly at the blue wing [-36; 9] km/s of the line. *Ninan et al. 2019* inferred column density of He($2^3S$) atoms at the level of $2.4 \cdot 10^{10}$ cm$^{-2}$. The simulation, based on 1D profiles of helium atoms and ions simulated in (*Salz et al. 2016*) assuming standard abundance He/H=0.1, yielded the column density one order of magnitude larger.

Very recently new three transits at the line of metastable helium have been reported by *Palle et al. 2020*. The obtained data has significantly better S/N ratio, better spectral resolution and better comparison of in-transit with out-of transit measurements. The spectra-photometric light curve has been obtained, which shows that the absorption in 10830 Å line coincides with the expected ingress and egress times. The spectrally resolved absorption at transit shows the depth of 1.5% around the line center restricted by the interval [-30; 20] km/s, while the half width interval of absorption is [-22; 10] km/s. Thus, the second measurement of HeI absorption more or less confirms the first one of (*Ninan et al. 2019*), albeit much more clearly.

*Palle et al. 2020* also simulated the absorption at 10830 Å line by 1D isothermal hydrodynamic model with non-equilibrium calculation of He($2^3$S) population. Similar to 1D modeling of *Ninan et al. 2019*, they found that He($2^3$S) distribution extends to at least 10R$_p$. The best fit to the observational data was achieved with the mass loss rate of $3 \cdot 10^{10}$ g/s at the thermosphere temperature of 6000 K and net blue shift of the line of 3.2 km/s. The line broadening was mainly due to radial expansion of gas with velocities up to 20 km/s.

These modeling results are reasonable first steps towards understanding of the nature of the observed absorption taking into account the uncertainty of many parameters involved and simplifying approximations assumed. So far, helium, supposed to be the second most abundant element of gaseous exoplanets, has been constrained by observations and modeling in a very few cases, HAT-P-11b (Allart et al. 2018), Wasp-107b (Spake et al. 2018), HD209458b (Lampon et al. 2020). The GJ 3470b seems to be different in that the absorption is likely produced rather far from the planet.

Therefore, in view of the available measurements, GJ 3470b is a challenging target for modeling by codes developed to simulate planetary wind (PW) of hot exoplanets and its interaction with the stellar environment. So far, such kind of modeling for this planet was done in *Bourrier et al. (2018)* using particle Monte Carlo code. The previous application of this approach to the similar system of GJ 436b has shown that the major factors which control the absorption in the blue wing of Lyα line are the stellar wind (SW), which deflects and disperses the atmospheric atoms escaping ahead of the planet, and the radiation pressure, which accelerates the hydrogen atoms (*Bourrier et al. 2015*, *Lavie et al. 2017*). Charge exchange of planetary atmospheric particles with the SW protons generates the hydrogen Energetic Neutral Atoms (ENAs) resulting in distinct ingress features in the spectral transit light curves of GJ 436b. The warm Neptune GJ 3470b is exposed to several times more intense Lyα flux, and it was found in *Bourrier et al. (2018)* that the stellar radiation pressure acting on the hydrogen atoms is strong enough to fit the observations, while any input from the SW protons is not required. At the same time, in order to explain the absorption in the red wing, an empirical exobase boundary of atmosphere, beyond which the Monte Carlo calculations were performed, was fitted in *Bourrier et al. (2018)* with a prescribed ellipsoid of a very large size, ~20R$_p$, allowing for sufficient spatial extension of the planetary atmospheric material in the direction of the orbital motion. It is obvious however, that the parameters and structure of dense plasmasphere around hot exoplanets interacting with the SW and extending far beyond the Roche lobe, which the Monte Carlo simulations take as the inner boundary, can be modeled only with the hydrodynamic codes.

In the present paper we use a 3D global hydrodynamic multifluid code which allows fully self-consistent calculation of the formation of PW and its interaction with the SW. Previously we employed this code to interpret the Lyα absorption at GJ 436b (*Khodachenko et al. 2019*) and the absorption in HI, OI, CII, SiIII resonant lines at hot Jupiter HD209458b (*Shaikhislamov et al. 2020*). For the GJ 3470b, we calculate the absorption both in Lyα and in He 10830 Å lines.

It should be noted that the modeling in such combination in 3D with the SW included is done for the first time. Similar to GJ 436b, we found that the absorption in the blue wing of Lyα line at transit of GJ 3470b can be explained only by the presence of resonant atoms (ENAs) formed due to charge exchange during the interaction between the huge and expanding planetary plasmasphere and the SW. At the same time, the contribution of properly included radiation pressure was found to be negligible, like in the cases of GJ 436b and HD 209458b. Our analytical estimates regarding the role of radiation pressure, supported also by the detailed numerical simulations, can be found in *Shaikhislamov et al.* (2016, 2018, 2020), *Khodachenko et al.* (2017, 2019), as well as in *Cherenkov et al. (2018)*.

The performed simulations predict also a moderate absorption in the red wing of the Lyα line when a bowshock is formed ahead the planet. In this case the thermal dispersion of the SW protons and that of the generated ENAs is large enough to generate sufficient thermal broadening of the Lyα line. In regard of He 10830 Å line, the calculation at parameters, for which the best fit for the Lyα data was obtained, revealed that in order to reproduce the absorption measured in *Ninan et al. 2019 and Palle et al. 2020*, the helium abundance in the upper atmosphere of GJ 3470b should be of about 7 times less than the typical Solar value.

The paper is organized as follows. Section 2 describes the model and the details of calculation of helium metastable $2^3$S state. In Section 3, the results of simulations are reported, followed by the discussion and conclusions in Section 4.

## 2. THE MODEL

The 3D multi-fluid hydrodynamic model used in the present work was developed on the base of previously employed 2D code (*Khodachenko et al. 2015, 2017, Shaikhislamov et al. 2016*) and has been already described in *Shaikhislamov et al.* (*2018, 2020*) and *Khodachenko et al.* (*2019*). Here we just repeat the most important details regarding the model. The code solves continuity, momentum, and energy equations for all species of the simulated multi-component flow. Among the considered species in this paper are the hydrogen and helium particles: H, $H^+$, $H_2$, $H_2^+$, and $H_3^+$, He, $He^+$, $He^{2+}$. The ENAs, generated by charge exchange between the planetary H and the stellar $H^+$, and having essentially different temperatures and velocities from those of H atoms of the planetary origin, are calculated as independent fluid. The hydrogen-helium photo-chemistry reactions are described in *Khodachenko et al.* (*2015*) and *Shaikhislamov et al. (2016)*. The hydrodynamic outflow is driven by heating by photo-electrons, which is derived by the integration of stellar XUV spectrum. As a proxy for the active M-dwarf GJ 3470, we use the modeled spectrum from *Bourrier et al.* (*2018*) averaged over 10 nm intervals. Being re-scaled to the reference distance of 1 a.u., the stellar radiation contains 4 erg s$^{-1}$ cm$^{-2}$ in the XUV band ($\lambda$<912 Å) and 3.5 erg s$^{-1}$ cm$^{-2}$ in the Ly$\alpha$ line. By this, the calculated photoionization time of hydrogen atom at the orbit of GJ 3470b is rather short, ≈1 hour.

The equations are solved in a so-called tidally locked planet based frame of reference on a spherical grid with polar-axis Z directed perpendicular to the orbital plane. In this frame we properly take into account the non-inertial terms, i.e., the generalized gravity potential and Coriolis force. To keep the number of points in the numerical code tractable for processing, the radial mesh is made highly non-uniform, with the grid step increasing linearly from the planet surface. This gives enough resolution in the highly stratified upper atmosphere of the planet, where the required grid step is as small as $\Delta r = R_p/400$, as well as in the inner regions of the expanding and escaping PW.

The SW plasma dynamics is calculated by the same code at the scale of the whole star-planet system, as described in *Khodachenko et al.* (*2019*) and *Shaikhislamov et al.* (*2020*). The SW is launched at the corona boundary and in the region R<20$R_{star}$ is accelerated by an empirical heating term derived from an analytical 1D polytropic Parker-like solution (*Keppens & Goedbloed 1999*). To correctly simulate shocks the equations are solved with the adiabatic index $\gamma=5/3$. Initial state is taken as a fully neutral atmosphere in a barometric equilibrium consisting from $H_2$ and He. At the planet boundary of the simulation domain at $r=R_p$ we fix zero velocity, the temperature 650 K, the pressure $p$=0.05 bar and the mixing ratio of He in relation to atomic hydrogen. To take into account the radiation pressure, we include the Ly$\alpha$ flux with the profile derived in *Bourrier et al.* (*2018*) and calculate its self-shielding in the ±30 km/s intervals.

| Reaction | Rate at T=$10^4$ K | Reference |
|---|---|---|
| He($2^3$S) → He($1^1$S) + h$\nu$ | 1.27·10$^{-4}$ s$^{-1}$ | Drake 1971 |
| He($2^3$S) + h$\nu$ → He$^+$ | 10$^{-3}$ s$^{-1}$ | Norcross 1971 |
| He($2^3$S) + e → He($2^1$S) | 2.68·10$^{-7}$ cm$^3$ s$^{-1}$ | Bray 2000 |
| He($2^3$S) + e → He($2^1$P) | 9.99·10$^{-8}$ cm$^3$ s$^{-1}$ | Bray 2000 |
| He($2^3$S) + e → He($3^3$S) | 2.06·10$^{-7}$ cm$^3$ s$^{-1}$ | Bray 2000 |
| He($1^1$S) + e → He($2^3$S) | 1.21·10$^{-8}$ cm$^3$ s$^{-1}$ | Bray 2000 |
| He($2^3$S)+H→ He+H$^+$ | 4.54·10$^{-9}$ cm$^3$ s$^{-1}$ | Roberge & Dalgarno 1982 |
| He($2^3$S)+H$_2$→ He+H+H$^+$ | 4.54·10$^{-9}$ cm$^3$ s$^{-1}$ | Roberge & Dalgarno 1982 |
| He$^+$(1S) + e → He($2^3$S) | 2·10$^{-13}$ cm$^3$ s$^{-1}$ | Osterbrock & Ferland 2006 |

**Table 1.** Reactions involved in populating the He($2^3$S) state.

Deriving the population of He atoms in the metastable triplet state ($2^3$S) involved in the absorption of 10830 Å line is a rather complex task (*Oklopcic & Hirata 2018, Ninan et al. 2019*). We simulate He($2^3$S) as a separate fluid, besides the ground state helium He($1^1$S), taking into account all the major excitation and de-excitation processes as listed in the Table 1. The reaction rates with electron impact in Table 1 are shown without a factor of exp(-$E_{ij}$/T)·$Y_{ij}$(T)/$Y_{ij}$($10^4$ K) where $E_{ij}$ is the gap between energy levels of a particular transition and $Y_{ij}$ is the corresponding collision strength. Around 75% of He$^+$ recombination contributes to the population of the He($2^3$S) state. The recombination rate depends on temperature as T$^{-0.75}$. Photoionization of He($2^3$S) is calculated using the energy dependent cross-section from Norcross (1971) integrated over GJ 3470 spectrum. The spectral range of 1215-2600 Å, required besides of XUV, we used from *Salz et al.* (*2016*).

The transition $2^3$P-$2^3$S of He is a triplet. The J=1 and J=2 lines are very close to each other and have the combined oscillator strength 8 times larger than J=0 line. Thus, we ignore the J=0 line and calculate combined absorption for J=1 and J=2 transitions only. The corresponding cross-section can be written as a sum of the *resonant*, or *thermal broadening* and *natural broadening*:

$$\sigma_{abs,He^*} (cm^2) \approx 1.2 \cdot 10^{-12} \cdot \sqrt{10^4 K/T} \cdot \exp(-x^2) + 9.4 \cdot 10^{-17} \cdot \left(\frac{10 \text{ km/s}}{V - V_z}\right)^2 \cdot q(x^2) \quad (1)$$

where $x = (V - V_z)/\sqrt{2kT/m_i}$, T, the temperature in Kelvin, $V_z$, the bulk velocity of absorbing particles along the line of sight (LOS), with a variable V, defining a particular point on the emission line profile expressed in terms of the Doppler shifted velocity. The used in Eq. (1) empirically parameterized function $q(x^2)$ was derived in (*Tasitsiomi 2013*). The analogous cross-section for the Lyα line was derived in *Shaikhislamov et al.* (*2018, 2020*) as follows:

$$\sigma_{abs,Lya} (cm^2) \approx 5.9 \cdot 10^{-14} \cdot \sqrt{10^4 K/T} \cdot \exp(-x^2) + 1.6 \cdot 10^{-17} \cdot \left(\frac{10 \text{ km/s}}{V - V_z}\right)^2 \cdot q(x^2) \quad (2)$$

## 3. RESULTS

The performed simulation runs demonstrate that at the reasonable parameters of the considered stellar-planetary system (see the parameter set No.1 in Table 2), a strong escaping PW is generated at GJ 3470b. However, under the conditions of weak SW, the created huge planetary plasmasphere does not result in the absorption values compatible with observations, neither in Lyα nor in He($2^3$S) line. This is because the size of transiting planet is rather small on the stellar disk, $(R_p/R_{Star})^2 \approx 0.0062$. Therefore, the natural line broadening in dense upper atmosphere close to the planet is insignificant, while for resonant absorption by rarified planetary material filling large volume around the planet, the velocity and temperature of PW are too low.

| No. | XUV | $M'_{pw}$ | $V_{sw,\infty}$ km/s | $T_{cor}$ MK | $M'_{sw}$ | $V_{sw,pl}$ km/s | $T_{swp}$ MK | $n_{swp}$ cm$^{-3}$ | $A_{Ly\alpha}$ blue, % | $A_{Ly\alpha}$ red, % | $A_{He}$ % | other |
|---|---|---|---|---|---|---|---|---|---|---|---|---|
| 1 | 4 | 2.6 | 360 | 3.5 | 7.5 | 200 | 0.65 | 7.5·10² | 8.0 | 5.4 | 2.0 | |
| 2 | 4 | 2.6 | 360 | 3.5 | 7.5 | 200 | 0.65 | 7.5·10² | 8.4 | 3.4 | | **rad. pressure** |
| 3 | 4 | 2.6 | 360 | 3.5 | **200** | 200 | 0.65 | 2·10⁴ | 22 | 5.6 | | |
| 4 | 4 | 2.6 | **720** | **5.0** | 62 | 520 | 1.05 | 2.5·10³ | 44 | 13 | 5.6 | |
| 5 | 4 | 2.2 | 720 | 5.0 | **45** | 520 | 1.05 | 1.8·10³ | 39 | 12 | 0.7 | **He/H=0.013** |
| 6 | 6 | 3.0 | 720 | 5.0 | 86 | 520 | 0.65 | 3.5·10³ | 36 | 6.6 | 0.7 | **He/H=0.013** |
| 7 | 6 | 3.0 | 720 | 5.0 | **170** | 520 | 0.65 | 7·10³ | 62 | 8.8 | 0.8 | **He/H=0.013** |
| 8 | **8** | 3.7 | 720 | 5.0 | 170 | 520 | 0.65 | 7·10³ | 39 | 5.9 | 0.7 | **He/H=0.013** |
| 9 | **10** | 4.5 | 720 | 5.0 | **340** | 520 | 0.65 | 1.4·10⁴ | 30 | 3.2 | 0.8 | **He/H=0.016** |
| 10 | **2** | 1.3 | 720 | 5.0 | 32 | 520 | 1.05 | 1.3·10³ | 76 | 16 | 0.8 | **He/H=0.013** |

Table 2. The modelling parameter sets used in the discussed simulation runs. XUV column lists the integrated stellar flux values in the range of wavelengths 10<λ<912 A, expressed in erg cm$^{-2}$ s$^{-1}$ and scaled to 1 a.u. The mass loss rates of the planet $M'_{pw}$ and the star $M'_{sw}$ are scaled in 10$^{10}$ g/s. Forth and fifth columns define Stellar Wind: terminal velocity and Coronal temperature. Index "swp" denotes the SW parameters at the orbit of the planet. The columns $A_{Ly\alpha}$, "blue" and "red", contain the absorption depths calculated at the mid-transit, and averaged over the blue [–94; –41] and red [23; 76] km/s wings of the Lyα line, respectively. Such averaging velocity intervals are chosen in order to compare the simulations with observations. The $A_{He}$ column shows the simulated absorption in the He 10830 Å (nm) line averaged over the interval [-40; 20] km/s. For the reference, the measured averaged values are 35% and 23% in the blue and red wing of the Lyα line, respectively, *from Bourrier et al. 2018*, 1% for He line from Ninan et al. 2019 and 0.87% from *Palle et al. 2020*. The last column contains the remarks on other modelling parameters varied additionally. If not specified otherwise, other parameters of the simulations are: $P_{base}$=0.05 bar, $T_{base}$=650 K, He/H=0.1. Bold letters indicate the parameters changed particularly for the respective run.

Figure 1 shows the color plots of the equatorial and meridional cuts of the simulated 3D spatial distribution of hydrogen atoms and protons around GJ3470b. The set of parameters (No.1) used for this particular model run, as well as those considered further on, are listed in Table 2. Note, that the mass loss rate is quite close to one calculated in *Salz et al.* (*2016*), and the difference by a factor of two can be explained by the twice smaller XUV flux used in our work.

Left panel in Figure 2 shows the simulated in-transit profiles of the Lyα line, as well as the averaged over the available observations out-of- and in-transit Lyα flux data points taken from *Bourrier et al.* (*2018*). To compare directly synthetic profiles with measurements we used intrinsic Lyα line profile absorbed by ISM from *Bourrier et al.* (*2018*), multiplied it by calculated transmission and convoluted with corresponding Line Spread Function of STIS. Two synthetic profiles are presented there – one obtained without the account of stellar radiation pressure (parameter set No.1) and another with the inclusion of the radiation pressure, assuming the corresponding integrated over the line Lyα flux of 3.65 erg cm$^{-2}$ s$^{-1}$ at 1 a.u. (parameter set No.2). Both profiles are close to the measured out-of-transit curve and show rather small absorption, with the depth averaged over the blue part of the line [-94; -41] km/s below 10%. Note that in the range of high velocities < −60 km/s the absorption is even less, being below 5%. In regard of red wing [23; 76] km/s of the Lyα line, absorption comes from the small velocities domain [23; 40] km/s, while at about 80 km/s it is only 3%.

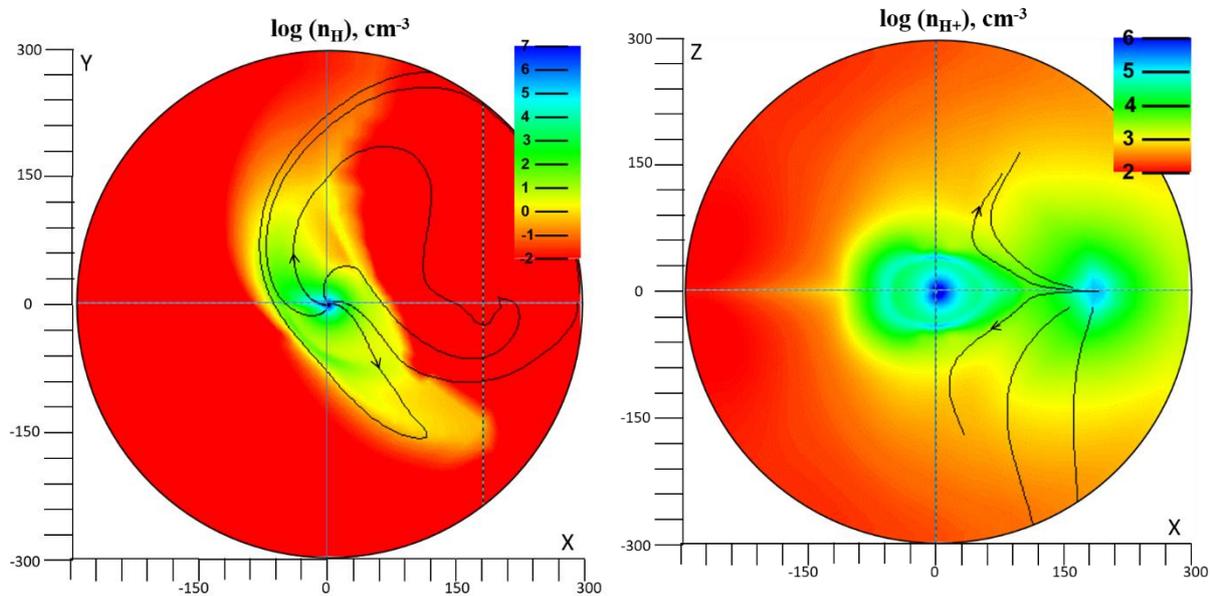

**Figure 1.** Color plots of atomic hydrogen (*left panel, equatorial plane*) and proton (*right panel, meridional plane*) densities, calculated with the set of parameters No.1. The distance is scaled in units of GJ3470b radius. The planet is located at the center of the coordinate system and moves anti-clockwise in the left panel; the star is to the right, at X=190. Black lines in both panels show the streamlines of the corresponding fluid.

As it can be seen from the comparison of plots in the left panel of Figure 2, the effect of radiation pressure is practically undistinguishable in the absorption. However, it is still rather important for the global dynamics, in particular, in the rarified region of the SW, preventing the hydrogen atoms from falling to the star. While without radiation pressure planetary material falls eventually to the star, as the streamlines in Figure 1 show, with radiation pressure included it is instead expelled to infinity. Also, the radiation pressure perceptibly decreases absorption in the red wing of Lyα line by about 2 % (see Table 2) due to decelerating the particles moving toward the star.

As we know from the modelling of other exoplanets (*Shaikhislamov et al. 2018, 2020, Khodachenko et al. 2017, 2019*), the asymmetry of absorption with the enhanced values in the blue wing of the Lyα line is a result of ENAs, generated due to the charge-exchange between the expanding PW atoms and the SW protons. Thus, as a reasonable way to reproduce the measured enhanced absorption in the blue wing of the Lyα line would be to enable interaction of PW with a relatively dense SW. The simulation run with the parameter set No.3 was performed to reproduce the SW conditions at the planetary orbit, similar to those of the slow Solar wind (terminal velocity 300-400 km/s, total mass loss ~2.5·10$^{12}$ g/s). As it can be seen in the Right panel of Figure 2, in this case a significant absorption, close to the measured values, takes place in the high-velocity blue-wing part of the Lyα line. Additionally to that, Figure 3 shows the simulated transit light-curves in the blue and red wings of the Lyα line, and compares them with the real measurements taken from *Bourrier et al.*

(*2018*). One can see that the transit in blue wing has an extended egress phase, not compatible with the observations. It is caused by the material, trailing behind the planet and extending along its orbital path, which can be clearly seen in Figure 1.

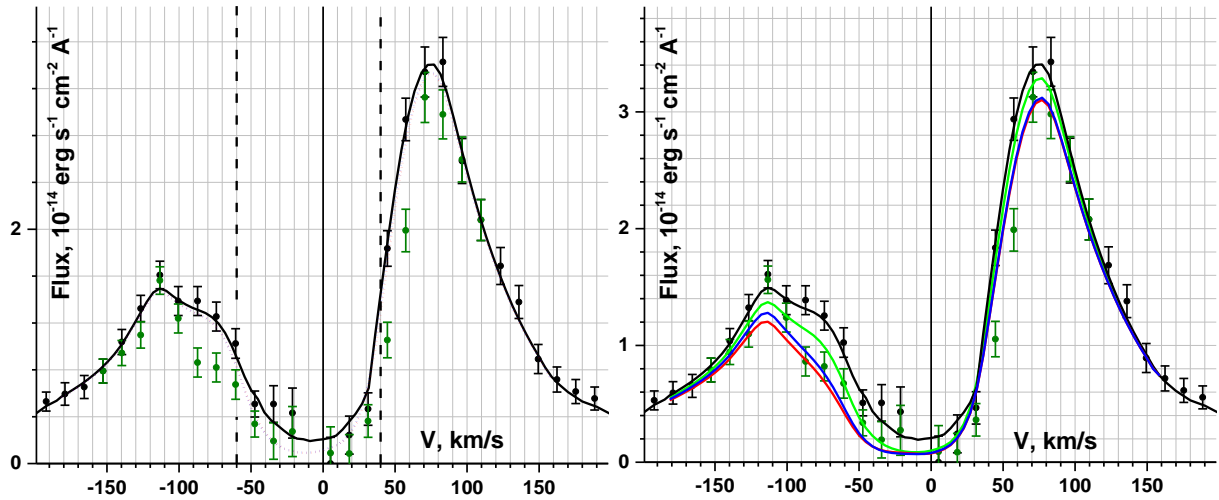

**Figure 2.** The absorption profiles of the Lyα line in Doppler shifted velocity units. Black and green dots with error-bars show the out-of-transit and mid-transit measurements, respectively, reproduced from *Bourrier et al. (2018)*. *Left panel:* The black line shows the line profile fit to the out of transit data, after convolution of intrinsic stellar line absorbed by ISM with the STIS LSF. The stellar line corrected by ISM absorption is taken from *Bourrier et al. (2018)*. The range of total ISM absorption, [-60; 40 km/s], is indicated by dashed vertical lines. Simulated mid-transit absorption profiles convoluted with STIS LSF are shown by red line (for parameter set No.1) and by blue line (set No.2). *Right panel:* Simulated mid-transit absorption profiles obtained with the parameter sets No.3 (green), No.4 (red), and No.5 (blue).

The similar effect was found also for GJ 436b (*Khodachenko et al. 2019*). The trailing tail can be blown away by a faster SW with, e.g., radial velocity significantly larger than the orbital velocity of planet, 116 km/s. To demonstrate this effect, the modelling run with the parameter set No.4 was performed with SW parameters analogous to fast Solar wind (terminal velocity ~700 km/s). The corresponding absorption profile/depth and transit light-curves in this case are relatively close to those measured in observations (Fig. 2 and 3).

It should be noted that the scenario with a slow SW (parameter set No.3) supposes a rather high density of SW, which is comparable in the mass loss to that of the Sun. This value may seem too large for an M-type dwarf star like GJ 3470. The fast SW (parameter set No.4) in contrast fits observations at about four times smaller stellar mass loss rate, which is compatible with the size of GJ3470.

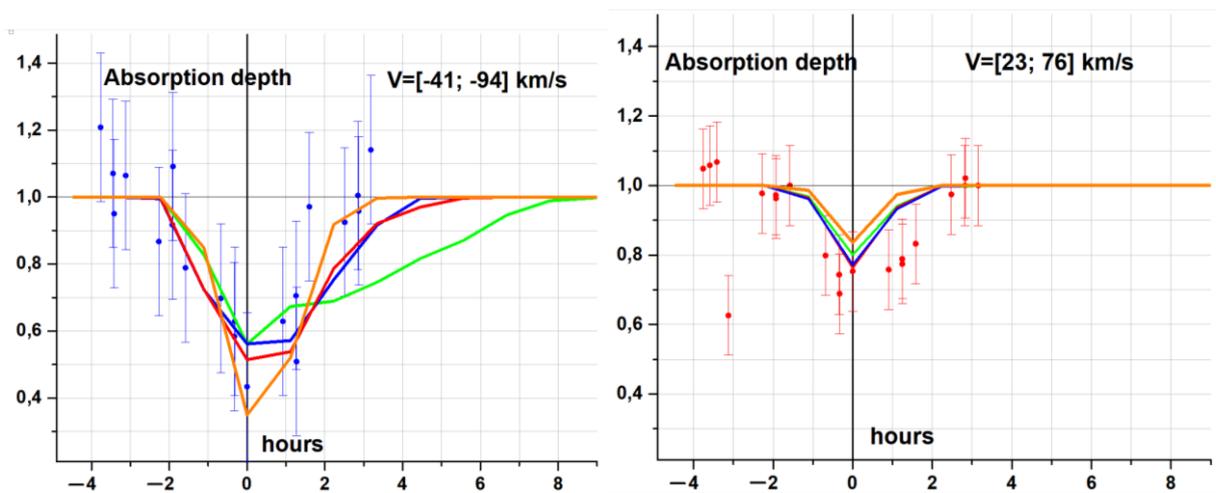

**Figure 3.** The simulated and measured transit light-curves of GJ3470b in the Lyα line averaged over the blue wing [-94; -41] km/s (*left panel*) and over the red wing [23; 76] km/s (*right panel*). Dots with error-bars show the measured values reproduced from *Bourrier et al (2018)*. The simulated transits were obtained with the parameter sets No.3 (green line), No.4 (red line), No.5 (blue line), and No.7 (orange line).

The particular difference of synthetic profiles from those measured in observations is that at red shifted velocities the simulated absorption averaged over the interval [23; 76] km/s is rather smaller than derived from the data by *Bourrier et al.* (*2018*) (~10% versus ~20%). Detailed comparison of out- and in-transit data in Fig. 2 shows that beyond the ISM total absorption window there is a range of moderate velocities [45; 70] km/s with large absorption of 24% and higher velocities [75; 85] km/s with moderate absorption of 10%. In these more restricted intervals the simulated absorption for sets No 4 and 5 is 13% and 9% respectively. We note that these values reflect the physics and not the effect of data processing as they are very close to the directly averaged transmissions calculated in the model at respective velocity intervals. Thus, while our modeling doesn't fully fit the red wing of the Lyα line at transit, it reproduces the absorption at particular high velocities.

To emulate these features *Bourrier et al.* (*2018*) used a very specific shape of the planetary plasmasphere, extending far ahead of the planet. In our opinion, the particulars of absorption in the red wing of Lyα can be due to the effect of magnetic field of SW, which directs a part of the PW flow along the planet-star line, reducing in this way the twisting of flow by the Coriolis force. Such structure of the PW and SW interaction was observed for the first time in 3D MHD simulation by *Zhilkin & Bisikalo* (*2019*).

The left panel of Figure 4 shows an equatorial cut of the simulated with the parameter set No.4, 3D distribution of ENAs generated in interaction of PW with the fast non-magnetized SW around GJ3470b. The Right panel in Figure 4 shows an equatorial cut of the corresponding 3D temperature distribution, which reveals the shape of a strong bowshock around the planet. Paying attention to the streamlines in the left panel of Figure 4, depicting the boundary of the bowshock, one can see that the ENAs are generated mostly inside the shocked region between the bowshock and ionopause. This feature has been also observed in our previous simulations of HD209458b and GJ346b (*Shaikhislamov et al. 2018, 2020, Khodachenko et al. 2017, 2019*). One can also clearly see in Figure 4 that the tail of escaping planetary material instead of trailing along the orbit is blown away by the strong and fast SW.

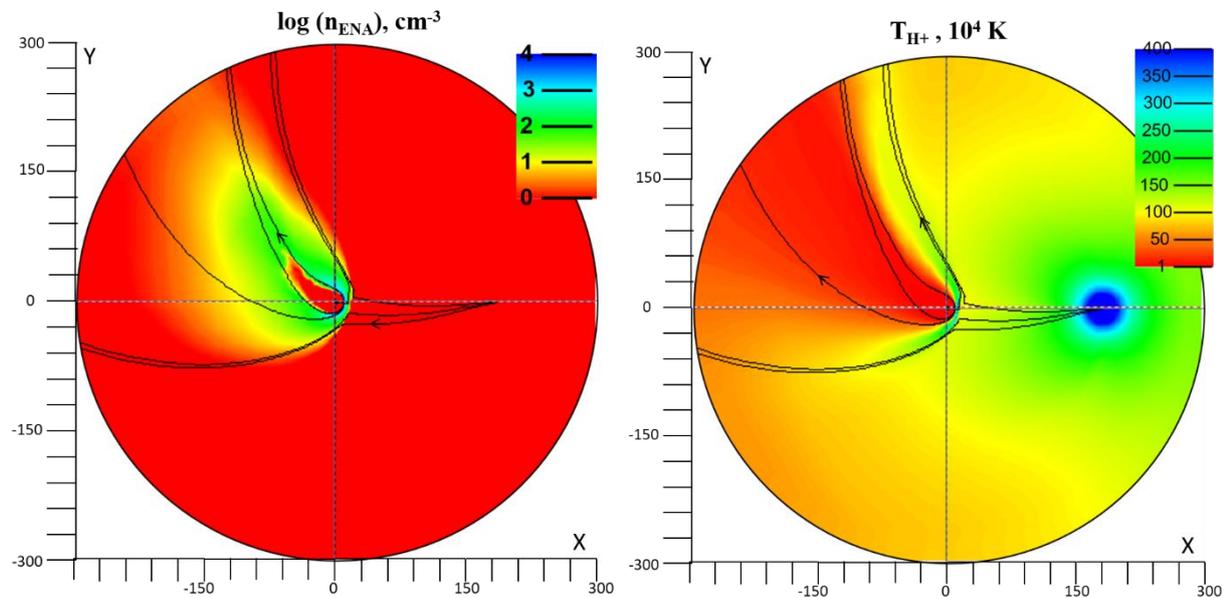

**Figure 4.** Color plots of the equatorial-plane distributions of the ENAs (*left panel*) and the proton temperature (*right panel*), simulated with the parameter set No.4. The planet is located at the center of the coordinate system, and the star is to the right, at X=190. Black lines in both panels show the streamlines of ENAs.

Next, let us discuss the simulations of absorption in He 10830 Å line. As described in Section 2, for this modelling we computed the population of He($2^3$S) state as a separate fluid according to the reactions' list in Table 1, along with the ground state populations of He, He$^+$ and He$^{2+}$. The obtained 3D spatial distribution of He($2^3$S) is then used to calculate the corresponding absorption across the stellar disk (see in Figure 5). The relative content of excited metastable helium atoms in $2^3$S state is orders of magnitude less than that of the ionized helium, so its contribution (addition) to the bulk effect is insignificant. The simulation run with the parameter set No.1, which corresponds to the case of a very week SW, and a typical for the Sun abundance of helium He/H=0.1, have shown an absorption with a maximum of up to 5% with an average value of 2.1%. This absorption takes place mostly within the velocity interval [-40; 40] km/s (see Figure 5, left panel, black dotted line). Thus, the absorption simulated without SW is larger than that measured in observations, and the maximum of the absorption profile is shifted to the red wing, while according to observation of *Ninan et al. (2019) and Palle et al. 2020,* the maximum of absorption is in the blue wing. It should be noted that synthetic profile is calculated at mid-transit, while observational data are averaged over the optical transit time. At that, spectra of *Palle et al. 2020* are correspondingly corrected to take into account the planet velocity change during transit, while that of *Ninan et al. 2019* are not corrected.

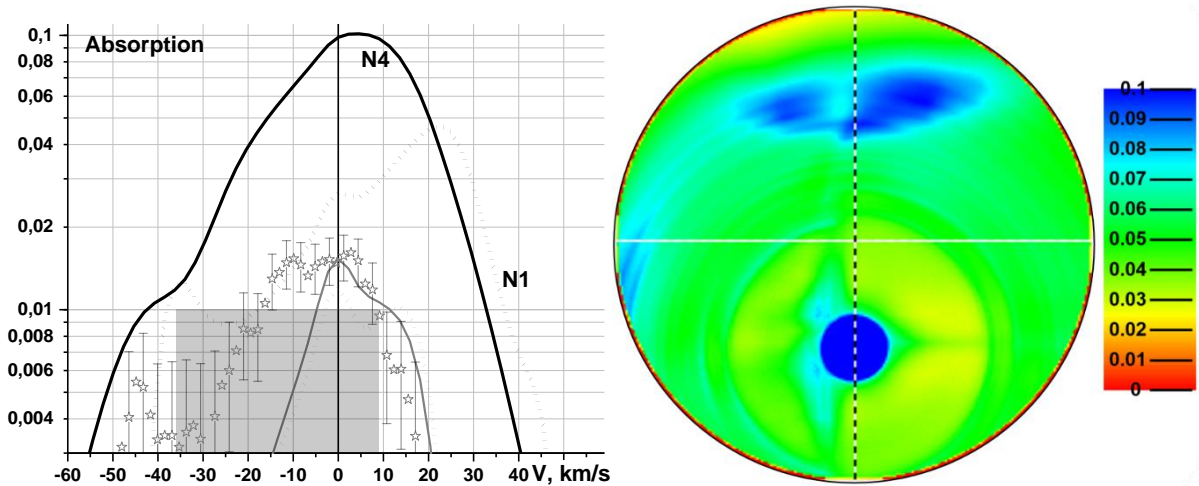

**Figure 5.** *Left panel*: The absorption profiles of He 10830 Å line in Doppler shifted velocity units, calculated with the parameter sets No.1 (dotted black line) and No.4 (solid black line) for the typical helium abundance He/H=0.1. The corresponding gray lines show the part of absorption produced at distances closer than 7 $R_p$ from the planet, i.e. within the region -7<X<7; $\sqrt{Z^2+Y^2}<7$. The shaded area represents the measurements by *Ninan et al. (2019)*. The stars with error bars represent the measurement by *Palle et al. (2020)*.
*Right panel*: Distribution of the absorption of He 10830 Å line across the stellar disk in the line interval [-25; 25] km/s, seen by an Earth-based observer at the mid-transit, simulated with the parameter set No.4. The filled blue circle indicates the planet.

The simulation run with the parameter set No.4, which shows good agreement with observations in Lyα line, reveals that the interaction of the escaping PW with the SW increases the absorption in He($2^3$P-$2^3$S) line, so it becomes significantly larger than that detected in *Ninan et al. (2019) and Palle et al. 2020* (maximum up to 10% with average value 4.6%, Fig. 5, left panel, black solid line). The absorption in the red wing of the line is produced mainly in the dayside part of the PW flow, which moves towards the star. The profiles in Figure 5 (left panel) show that absorption takes place in the line core and at the relatively far wings. An interesting point in that respect is that the line core absorption is not produced in the region close to the planet, as it usually happens for the majority of lines. To demonstrate this, we calculated the absorption produced by the material within a region of $7R_p$ around the planet, and found that it reaches only 1%. It is worth to note, that the SW does not affect the absorption within this region, as one may expect, but it significantly adds to the absorption produced beyond $7R_p$. Thus, the material, moving relatively far from the planet provides the major contribution to the absorption in the whole He line. This is clearly demonstrated in Figure 5 (Right panel), which shows the distribution of absorption over the stellar disk, as it is seen by an Earth-based remote observer. It is more-or-less smoothly distributed over the stellar disk with the maximum of absorption produced in the shock region, relatively far from the planet.

To understand the process of populating the metastable $2^3$S state of helium, we plot in Figure 6 the distributions of species, involved in the key reactions listed in the Table 1, along the axis crossing the bowshock ahead of the planet shown in Figure 4. One can see that the density of metastable He($2^3$S) has two maxima comparable in value, one is close to the planet, and another appears in the shocked region between the ionopause and bowshock. Right panel in Figure 6 reveals that inside the PW dominated region the He($2^3$S) level is pumped by recombination of He$^+$. However, in the low atmosphere where the density of He$^+$ is at maximum, it is rapidly depopulated via auto-ionization reaction with $H_2$ and H so, that the density of He($2^3$S) remains relatively low to make any significant contribution to the absorption. Thus, the absorption comes from an extended region around the planet where $H_2$, H densities are relatively low and the recombination pumping is balanced mostly by electron impact de-excitation.

In this region the velocity of PW is already significant, and the effects of advection and flow divergence become important in the formation of density distribution of species. Note that the photo-ionization and the radiative decay of He($2^3$S) remain insignificant here. The surprising finding is that in the case when the interaction between the PW and SW is properly taken into account, the shocked region becomes the largest contributor of the broad band He absorption. The temperature there is more than enough to pump the He($2^3$S) from the ground state by direct electron impact. The helium atoms, being neutral, are able to penetrate into the shocked region across the ionopause, while ions are mostly stopped there. The corresponding collisional mean free path is large in the shocked region. Therefore, the neutral helium particles are mostly uncoupled from other species and fly almost ballistically until ionized by photons or hot electrons. This is to certain extend similar to the process of generation of ENAs, when the hydrogen atoms penetrate into the shocked region and interact there with the SW particles (*Shaikhislamov et al. 2016, 2018, 2020, Khodachenko et al. 2017*). Note, that the Lyα absorption simulated in the present modelling of GJ 3470b is produced by ENAs generated in such a way. However, while velocity of ENAs comes from the energetic stellar protons, the velocity and temperature distributions of the absorbing He($2^3$S) particles come from the helium atoms. In the shocked region they are quite different from those of ions. This is the reason why the absorption in He($2^3$S) line, in spite of being produced in the simulation run No.4 mostly in the shocked region, is relatively symmetric over the blue and red wings of the line. Note for comparison, that in the case of Lyα absorption, the latter is significantly shifted to the blue part of the line because of the stellar proton velocities, dominating among the absorbing ENAs. This explanation was carefully checked with the detailed comparison of the velocity streamlines of helium atoms and ions.

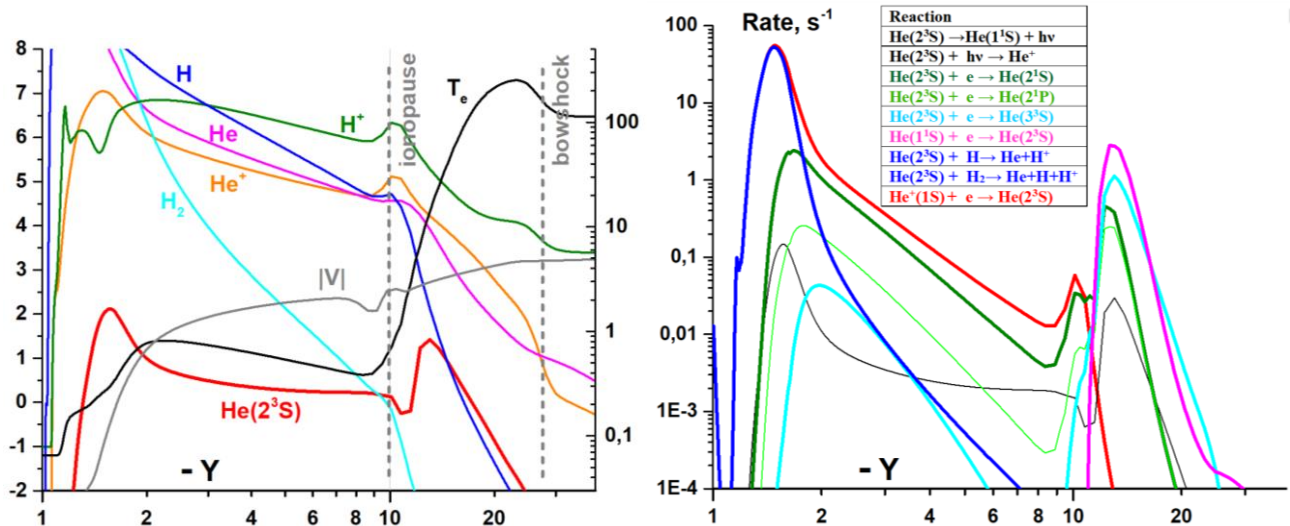

**Figure 6.** The profiles of major material components along the negative direction of Y axis.
*Left panel, left axis*: log of density of atomic hydrogen (blue), molecular hydrogen (cyan), protons (green), helium ions (orange), helium atoms in ground (magenta) and metastable states (red), respectively.
*Left panel, right axis*: electron temperature (black, in units of $10^4$ K) and velocity of He($2^3$S) particles (gray, in units of 10 km/s). Vertical dashed gray lines indicate the positions of the bowshock and ionopause, respectively.
*Right panel*: Rates of reactions from Table 1 involved in the production of He($2^3$S) particles. Black line shows the sum of reactions 1 and 2, and blue line, the sum of reactions 7 and 8.

To achieve the simulated absorption level in the He 10830 Å line comparable to observations, i.e., ~ 1%, the helium abundance should be reduced by 7.5 times, as compared to the initially used Solar value, down to He/H=0.013. The performed modelling also reveals that the only way to shift the absorption maximum to the blue part of the line is to increase the contribution of the shocked region and to increase the shock pressure. Another factor, which influences the absorption in He 10830 Å line is the XUV radiation flux. For example, the increase of XUV flux up to 8 erg s$^{-1}$ cm$^{-2}$ results in a faster photoionization. Therefore, less helium atoms are able to reach the shocked region, and hence, less absorption in the corresponding He line will be produced there. However, without broadband absorption in the shocked region, the total absorption is shifted to the red part of the line, like in the simulations with the parameter set No.1, i.e., without SW.

Figure 7 shows the absorption profiles of He($2^3$S) and Lyα lines simulated with different modelling parameter sets (No.6, No.7, No.8, No.9), assuming in all cases low values of the helium abundance, as specified in Table 2. The parameter set No.9, with a bit higher helium abundance, He/H=0.016, denser SW and essentially higher XUV flux, provides the best fit of the blue shifted absorption in both, He($2^3$S) and Lyα lines. However, the measured values of the Lyα absorption in the red wing of the line still cannot be reproduced at the same time. The parameter set No.7, with the moderate SW density, XUV flux and lesser helium abundance, He/H=0.013, enables obtaining of a reasonable blue shifted absorption in the He 10830 Å line and the perceptible absorption values in the Lyα line at high red-shifted velocities. At the same time, the absorption in the blue wing of the Lyα line, simulated with the parameter set No.7, is rather high. The transit light-curves, obtained with the parameter set No.7 in blue and red wings of the Lyα line, are shown in Figure 3. A simulation run with the low XUV flux, assumed in the parameter set No.10, gives the absorption profile of He($2^3$S) line almost identical to the case of parameter set No.9, which has the maximal XUV flux and just slightly higher helium abundancy. However, the averaged absorption in Lyα line is very different for these two cases (see Table 2).

To check how the decrease of helium photoionization affects the absorption in He($2^3$S) 10830 Å line, we performed several trial runs with the same parameter sets as in Table 2, but with the 10 times attenuated XUV flux at λ<504 Å, as was considered in *Ninan et al. (2019)*. Opposite to the qualitative modeling of *Ninan et al. (2019)*, it was found that in all cases the absorption increases. For the parameter set No.1 it becomes fully red-shifted, while for No.4 it remains similar in shape to the calculation without flux attenuation, but is two times larger.

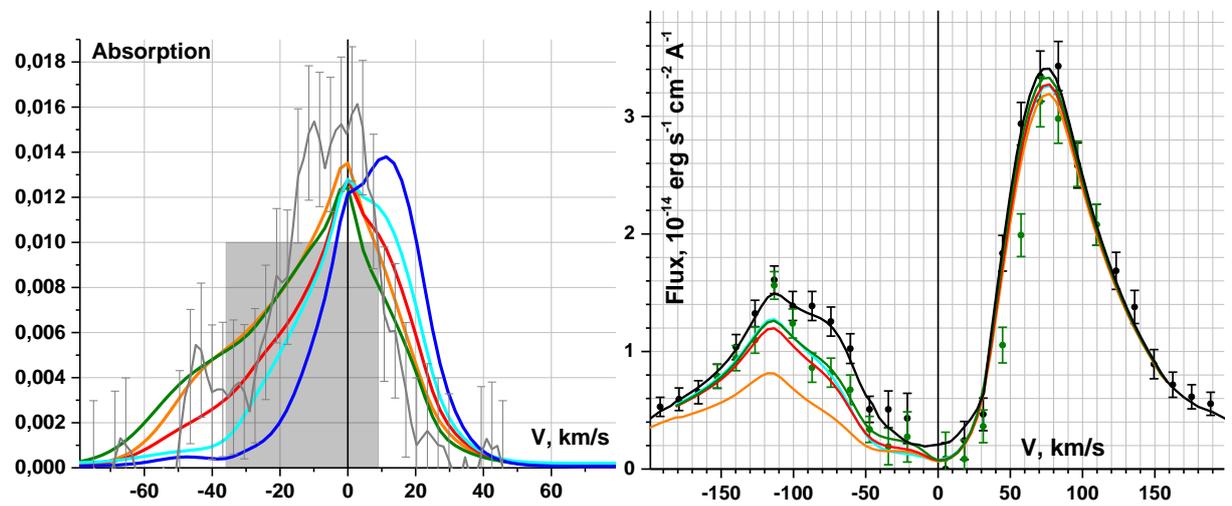

**Figure 7.** *Left panel*: The absorption profiles of He 10830 Å line in Doppler shifted velocity units, calculated with the parameter sets No.5 (blue line), No.6 (cyan), No.7 (orange), No.8 (red), No.9 (green) for the low helium abundances, specified in Table 2. The shaded area represents the measurements by *Ninan et al. (2019)*. The grey line with error bars represent the measurement by *Palle et al. (2020)*.
*Right panel*: The corresponding absorption profiles of Lyα line simulated with the parameter sets No.6, No.7, No.8, No.9. The lines color coding is the same as in left panel. Black and green dots with error-bars show the out-of-transit and mid-transit measurements, respectively, reproduced from *Bourrier et al. (2018)*.

## 4. DISCUSSION AND CONCLUSION

The warm Neptune GJ3470b offers, along with the GJ436b, a good target for the extensive quantitative application of numerical models of the expanding upper atmospheres of hot close-orbit exoplanets. Moreover, its special uniqueness in the present time consists in the availability of high quality measurements of the planet transit in Lyα, combined with the detection of absorption in metastable helium He($2^3$S) line. The notable difference between GJ3470b and GJ436b concerns the relatively short egress and significant absorption in the red wing of the Lyα line demonstrated by GJ3470b. The application of global 3D HD multi-fluid model, reported in this paper, has shown that, similar to the GJ436b (*Khodachenko et al. 2019*), the Lyα transit for GJ3470b is produced by ENAs formed in course of the interaction between the escaping PW and the surrounding SW. In contrast to hot Jupiter HD209458b (*Shaikhislamov et al. 2018, 2020, Khodachenko et al. 2017*), the absorption due to natural line broadening is insignificant for both warm Neptunes, because of the relatively small planets' size. The role of the radiation pressure acceleration in the case of GJ3470b has been shown to be insignificant, like in our previously reported studies of HD20458b and GJ436b (*Shaikhislamov et al. 2018, 2020, Khodachenko et al. 2017, 2019*). This result opposes the conclusion of *Bourrier et al. (2018)* based on Monte-Carlo modeling. The applicability of either fluid or particle approach is not fully justifiable to the considered problem, and the conclusion on relative roles of the radiation pressure and the SW-PW interaction in producing of ENAs is still debated.

In course of the performed simulation runs with different modelling parameter sets we have found that fast decay of the Lyα absorption in the egress part of the transit light-curve can be the result of sufficiently high pressure of fast SW, which rapidly blows away from the orbit the trailing tail of escaping planetary atmospheric material. The applied model reproduced, within the measurement error bar, the 10% Lyα absorption at red-shifted velocities of about 80 km/s. The red-shifted Lyα absorption is generated in the shocked region by ENAs, which have sufficiently high velocity dispersion to provide absorption in the blue and red parts of the line. However, it should be acknowledged, that the model doesn't fit details of absorption in the red wing, especially at velocity range [45; 70] km/s.

The best fit parameters of the SW correspond in the temperature and velocity to the fast Solar wind and imply for the parent star of GJ3470b the total mass loss rate of ~0.1 of the Solar value, which is compatible with the smaller than the Sun size of the star. The simulation of Lyα transits for GJ436b and GJ3470b with our multi-fluid self-consistent hydrodynamic modeling reveals that both cases require the SW plasma parameters not very much different from those taking place in the Solar wind. Moreover, the absorption depth in the blue wing of the Lyα line is directly proportional to the SW density, whereas the absorption velocity range – to the SW relative velocity at the planet orbit. This fact allows a quantitative constraining for the SW parameters based on observations.

Besides of Lyα, the absorption by the metastable $2^3S$ level of helium is self-consistently modelled in this paper for the first time. So far, this kind of studies was attempted only with non-self-consistent models, using as an input the results derived from a 1D modeling. Despite of the restrictions of 1D geometry and other limitations of the previous studies, an averaged over the whole stellar disk column density of He($2^3S$), calculated e.g., in *Ninan et al. (2019)* on the basis of 1D simulation by *Salz et al. (2016)*, appears to be very close to the value NL=$3.1 \cdot 10^{11}$ cm$^{-2}$, obtained by our model with the parameter set No.4.

Measurements of *Palle et al. 2020* give an opportunity of more detailed comparison with modeling. It should be noted that we have not performed simulations to fit these measurements because the paper of *Palle et al. 2020* was still not released at the time. However, one can see from Fig. 5 that there is reasonably good agreement. For the run No.8 the absorption at maximum and in the blue wing can be fitted more closely to the measurements by increasing the Helium abundance by 1.25 times to He/H=0.016. However, like in the case of comparison with the data from *Ninan et al. (2019)*, such fitting will put the absorption in the red wing above the measurement level. Introducing the empirical net blue shift of about 3-5 km/s, similar to modeling of *Palle et al. 2020*, will make the synthetic curve to fit the measurements within the error bars.

There is also a good agreement with 1D modeling of *Palle et al. 2020* regarding the mass loss rate ($3 \cdot 10^{10}$ g/s) and velocities of planetary outflow (5-20 km/s), derived from fitting the He($2^3S$) absorption, which well correspond to obtained in this paper. Thus, there is good general agreement between different aeronomy models and modeling approaches. The main conclusions followed from all models, is that the He($2^3S$) particles are distributed in a wide area around the planet ~10$R_p$. Our finding that in order to match the absorption at a level of 1%, the calculated population of He($2^3S$) should be reduced by about an order of magnitude, if the standard abundance He/H=0.1 is assumed, repeats the conclusion of *Ninan et al. (2019)*. However, consequent and accurate account of the effects of stellar XUV radiation and SW, as well as the geometry and structure of the material flows to simulate a realistic line absorption profiles and transit light-curves is possible only in a 3D model. For example, significant reduction of XUV flux below λ<504 Å proposed in *Ninan et al. (2019)* as a possible way to fit observations, in our self-consistent model leads to an opposite effect, especially if the interaction between the escaping PW and the surrounding SW is properly described and taken into account. The reduction of helium photoionization, caused by the lower XUV flux, reduces the number of He$^+$ ions and the related with them recombination pumping of the He($2^3S$) state. At the same time, it increases the number of neutral helium atoms, penetrating into the shocked region, where He($2^3S$) state is pumped by the electron impact. The comparative study of various pumping and depopulation processes of He($2^3S$) state, performed in this paper, has shown that the interpretation of the measured absorption in He($2^3S$) 10830 Å line can be quite intricate, because in different regions of the escaping PW, as well as in the shocked region, different processes are responsible for the production of absorbing agent. It should be also noted, that the most of the reaction rates, used in the modeling of the metastable helium state $2^3S$, were calculated theoretically and without substantial experimental verifications. Therefore, these rates might be reconsidered in future and affect the simulation results.

Altogether the performed 3D self-consistent multi-fluid simulations of the expanding and escaping upper atmosphere of GJ3470b and the related spectral absorption features and transit light-curves have shown that the available observational data can be relatively well interpreted within the range of reasonable values of physically justifiable parameters related with the planetary atmospheric composition, stellar XUV flux and SW plasma flow. However, to quantitatively constrain these parameters and obtain better agreement more observations and modelling are needed, with the inclusion of important factors and processes, which were not accounted in the models so far.

**Acknowledgements:**
This work was supported by grant № 18-12-00080 of the Russian Science Foundation. Parallel computing simulations, key for this study, have been performed at Computation Center of Novosibirsk State University, SB RAS Siberian Supercomputer Center, Joint Supercomputer Center of RAS and Supercomputing Center of the Lomonosov Moscow State University.